\documentclass[aps,pre,floatfix,twocolumn,a4paper,showpacs]{revtex4}
\usepackage{amsmath}
\usepackage{amsfonts}
\usepackage{graphicx}
\usepackage{epsfig}
\usepackage{color}
\usepackage{xspace}
\usepackage{subfigure}

 \newcommand{\ket}[1]{ \ensuremath{\left| #1 \right\rangle} }

 \newcommand{\EX}[1] { \ensuremath{\left\langle #1 \right\rangle} }

 \newcommand{\half} {\ensuremath{\frac{1}{2}}}
 \newcommand{\ihbar}{\ensuremath{\frac{i}{\hbar}}}

 \newcommand{\BUE}{Centre for Theoretical Physics,
                   The British University in Egypt,
                   El Sherouk City, Misr  Ismalia Desert Road,
                   Postal No. 11837, P.O. Box 43, Egypt.}

\begin{document}
\title{Recovery of classical chaotic-like behaviour in a conservative quantum three body problem}
\author{M.J.~Everitt}
\email{m.j.everitt@physics.org}
\affiliation{\BUE}

\begin{abstract}
Recovering trajectories of quantum systems whose classical counterparts display chaotic behaviour has been a subject that has received a lot of interest over the last decade. However, most of these studies have focused on driven and dissipative systems.
The relevance and impact of chaotic-like phenomena to quantum systems has been highlighted in recent studies which have shown that quantum chaos is significant in some aspects of quantum computation and information processing.
In this paper we study a three body system comprising of identical particles arranged so that the system's classical trajectories exhibit Hamiltonian chaos. Here we show that it is possible to recover very nearly classical-like, conservative, chaotic trajectories from such a system through an unravelling of the master equation.
Firstly, this is done through continuous measurement of the position of each system. Secondly, and perhaps somewhat surprisingly, we demonstrate that we still obtain a very good match between the classical and quantum dynamics by weakly measuring the position of only one of the oscillators.
\end{abstract}
\pacs{05.45.Mt 03.65.-w 05.45.Pq}
\maketitle

\section{Introduction}

Quantum mechanics is perhaps the most powerful and useful theory of physics to date. Indeed, with the possible emergence of many new quantum technologies in areas such as computation, communication, cryptography and metrology this trend looks set to continue well into the future. With such strong interest in applications of quantum mechanics comes a concomitant interest in the measurement process and the interaction between \emph{``classical''} and quantum systems. Indeed, as we wish to understand and apply quantum mechanics within the context of modern technology we will need to develop our understanding of what actually constitutes a classical device and how such objects interact with  quantum systems. However, the recovery of classical mechanics is not always as simple as a
implied by a \emph{na\"ivet\'e}  interpretation of the correspondence principle. This essential requirement of any physical theory can, for quantum mechanical systems, be stated as:
\begin{quote}
        \emph{``If  a quantum system  has a  classical analogue,  expectation values  of 
        operators behave, in  the limit  $\hbar\rightarrow 0$, like  the corresponding
        classical  quantities''}  \cite{Mer98}
\end{quote}
We observe that interpretation of this statement can be problematic if, for example, we consider quantum systems that lack a specific dependence on Planks constant~\cite{Cas79}. Further difficulties arise when attempting to recover the classical trajectories of classically non-linear and chaotic systems as the Schr\"odinger equation is strictly linear. 

We note that these concerns are no longer just of interest to those studying either the measurement problem or the correspondence principle and the emergence of the classical world. Indeed this
area has a direct impact on quantum technologies. In order to fully leverage the power afforded by these emerging fields we must not only understand in depth the measurement process but also many body quantum systems coupled to real environments. This is highlighted by the recent observation of chaos in the spectrum
of Shor's algorithm~\cite{maity06} as well as in other studies involving quantum information processing and quantum chaos~\cite{lages06,kiss06,Rossini06}.

A solution to the correspondence problem for chaotic systems which has been employed with great success is found by utilising quantum trajectories methods~\cite{Car93,Gis93,Gis93b,Heg93,Wis96,Ple98,spi1,bru1,bru2,zur1,sch1}. Here, introduction of environmental degrees of freedom and unravelling the master equation yield stochastic Schr\"odinger equations from which chaotic-like trajectories may be recovered. This process can be considered as comprising of several steps. Firstly, we make the quantum system of an open one. This is archived by coupling the quantum object to an environment which may take the form of a measurement device. Once the environment has been introduced we model the evolution of the system's density operator (in the presence of the environment) using a linear master equation. However, master equations are similar to the Langevin equation insofar as they only predict a set of probable outcomes over an ensemble of systems or experiments. Therefore, in order to get some idea of the possible behaviour of an individual experiment we next unravel the master equation. In essence, this process involves finding a stochastic differential equation for the system's state vector with the proviso that the dynamics given by the master equation are returned in the ensemble average over many solutions. There are an infinitely many ways to do this each representing a different physical process. In this work we employ the quantum state diffusion (QSD) unravelling which corresponds to a unit-efficiency heterodyne measurement (or ambi-quadrature  homodyne detection) on  the environmental degrees of freedom~\cite{Wis96} (for an detailed introduction to this approach please see~\cite{Per98}). Here the evolution of the  state
vector $\left| \psi\right\rangle$ is given
by the It\^{o} increment equation~\cite{Gis93,Gis93b}
 \begin{eqnarray}\label{eq:qsd}
 \ket{d\psi}  &  =&-\ihbar H \ket{\psi} dt\nonumber\\
 &&  +\sum_{j}\left[  \EX{L_{j}^{\dagger}} L_{j}-\half L_{j}^{\dagger}L_{j}-\half \EX{L_{j}^{\dagger}} \Bigl\langle L_{j}\Bigr\rangle \right]  \ket{\psi} dt\nonumber\\
 &&  +\sum_{j}\left[  L_{j}-\EX{L_{j}} \right]  \ket{\psi} d\xi
 \end{eqnarray}
where  the  Lindblad  operators   $L_{i}$  represent  coupling  to
environmental  degrees of  freedom, $dt$  is the  time  increment, and
$d\xi$     are     complex     Weiner     increments defined    such     that
$\overline{d\xi^2}=\overline{d\xi}=0$        and       $\overline{d\xi
  d\xi^{*}}=dt$~\cite{Gis93,Gis93b}. Throughout this work a bar over a quantity denotes the average over stochastic processes whilst the notation $\EX{\cdot}$ is used for quantum mechanical expectation values. The first term on the right hand side of this equation is  the Schr\"odinger evolution
of the system  while the second (drift) and  third (fluctuation) terms
describe the decohering effects of the environment on the evolution of
the systems state vector. 

However, to date the body of work which uses quantum trajectories to recover classically chaotic-like trajectories has focused on those systems that are dissipative. 
There are several notable exceptions that demonstrate that continuous measurement of both driven
and un-driven conservative systems that can recover classical like behaviour~\cite{bha1,bha2,sco1,gho1}. However, these works consider systems with only a single degree of freedom.
Recently we became interested in whether it was possible,  using a similar analysis, to recover chaotic trajectories of classical, multi-component, systems undergoing Hamiltonian chaos. Initially we wished to consider the traditional three body problem of classical mechanics for particles with similar masses~\footnote{We note that although the helium atom is also a standard example of a three body system, due to the very large mass of the nucleus, this particular example simplifies to the restricted three body problem}. This problem, although historically very significant, is non-trivial to solve. Consequently, in this work we consider a  somewhat simplified system comprising of three coupled one-dimensional anharmonic oscillators.

\section{Background}
The Hamiltonian for our chosen three body system, comprising of one-dimensional anharmonic oscillators with a quartic potential and unit mass, is given by
\begin{equation}
H = \frac{1}{2}\sum_{i=1}^3 p_i^2+\beta^2\left(
\frac{q_1^2q_2^2+q_2^2q_3^2+q_1^2q_3^2}{2}+
\frac{q_1^4+q_2^4+q_3^4}{32}
\right)\label{eq:Ham}
\end{equation}
The classical dynamics associated with this Hamiltonian can be chaotic and are known to have positive Lyapunov exponents~\cite{rugh97}.
When we consider classical mechanics $q_i$ and $p_i$ are taken to represent the classical values of position and momentum. However, when we consider the quantum mechanics they are replaced by their operator counterparts. As we shall always be clear as to which description we are considering at any one time this does not lead to any ambiguity.

We have already stated one expression of the correspondence principal in quantum mechanics. An alternative definition, which we find preferable, is to 
\begin{quote}
\emph{``consider $\hbar$ fixed (it is) and scale the Hamiltonian so  that  the  relative  motion  of  the  expectation  values  of  the observable   become   large   when compared   with   the   minimum   area $\left(\hbar/2\right)$ in the phase  space''}.
\end{quote}
 In either case this is the \textit{r\^ole} of the $\beta$ term in the Hamiltonian, i.e. $\hbar \rightarrow \beta \hbar$ so that the smaller $\beta$ the larger the dynamics when compared to a plank cell.

\section{Results}
\begin{figure}[!t]
\begin{center}
\resizebox*{0.48\textwidth}{!}{\includegraphics{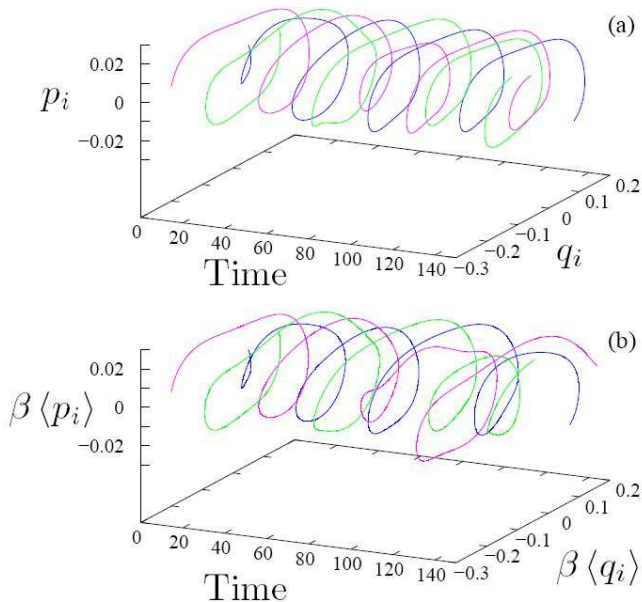}}
\caption{(colour on-line) An example chaotic trajectory (a) solutions to the classical equations of motion~(\ref{eq:cls}) for the initial conditions $q_1=-0.2$, $q_2=0.05$, $q_3=0.15$ for our chosen system and (b) scaled quantum expectation values $\beta \EX{q_i}$ and $\beta \EX{p_i}$ versus time for an unravelling of the master equation with initial state $D_1(-0.2/\beta)D_2(0.05/\beta)D_3(-0.15/\beta)\ket{000}$ and $\beta=1/2000$. Trajectories for oscillator one are shown in magenta (medium grey), for two in blue (dark grey) and three in green (light grey). Note, all quantities are dimensionless.
\label{fig:1}}
\end{center}
\end{figure}

From the Hamiltonian~(\ref{eq:Ham}) we find the three classical equations of motion are
\begin{eqnarray} \label{eq:cls}
\ddot{q_i}+\beta^2 \left\{\frac{q_i^3}{8} +\sum_{i \neq j} q_i q_j^2\right\}=0,\ \mathrm{where}\ i,j=1,2,3.
\end{eqnarray}
When we solve these coupled equations of motion with the initial conditions $q_1=-0.2/\beta,\ q_2=0.05/\beta,\ q_3=0.15/\beta$ and $p_i=0$ for all $i$ we find that the dynamics are chaotic. We show the phase portrait for the solutions to these equations in figure~\ref{fig:1}(a) where, without loss of generality, we have set $\beta=1$.
We note as an interesting aside, one feature of this system is that if $q_i=p_i=0$ for any $i$ then $q_i=p_i=0$ always.

\begin{figure}[!t]
\begin{center}
\resizebox*{0.48\textwidth}{!}{\includegraphics{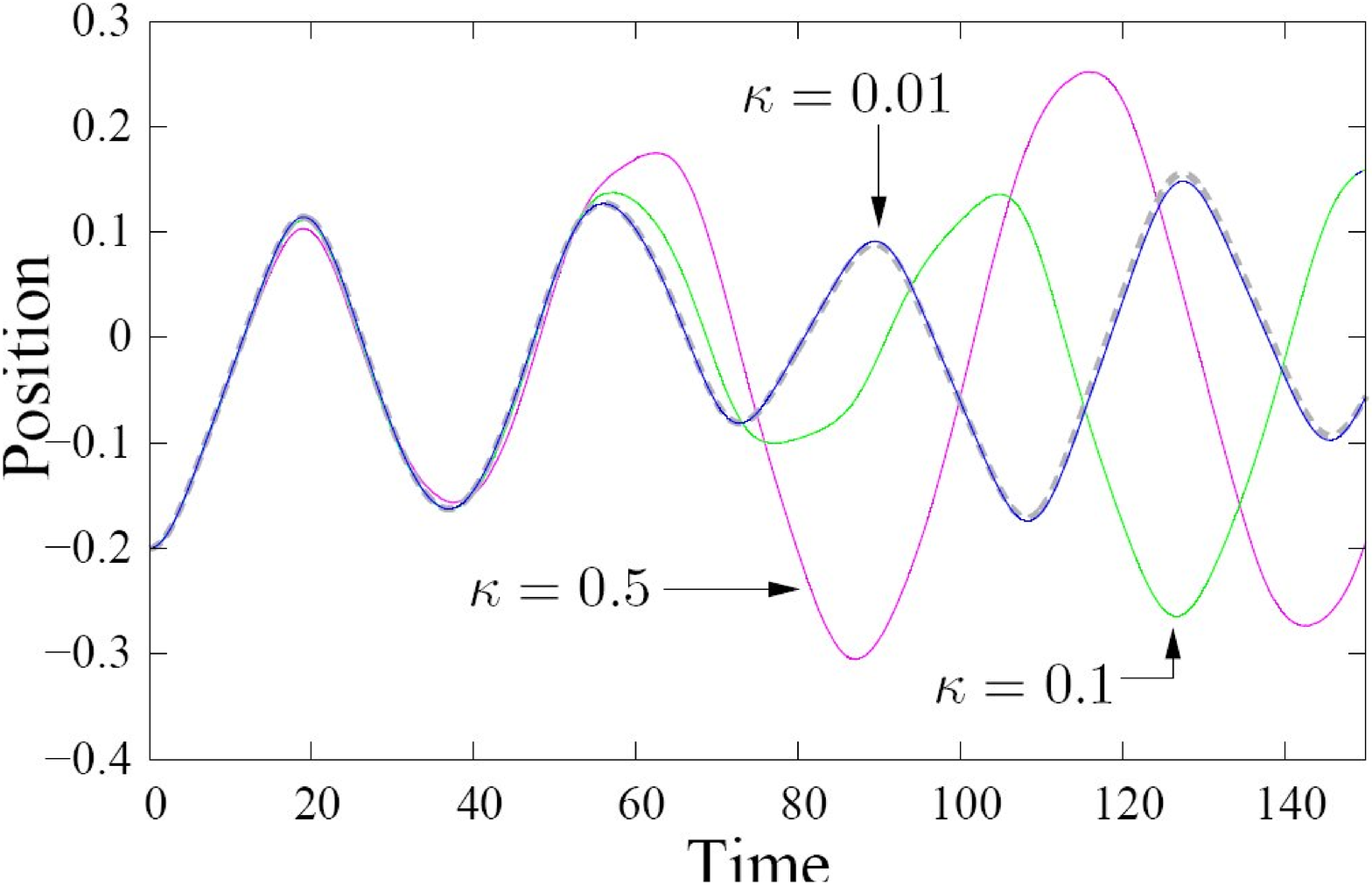}}
\caption{(colour on-line) A comparison between the classical position $q_1$ (dashed grey) and  $\beta\EX{q_1}$ for three different couplings ($\kappa$) to the environment. Note, all quantities are dimensionless.
\label{fig:new2a}}
\end{center}
\end{figure}

We now proceed to discuss the quantum mechanical description of  these coupled oscillators. Unlike the classical equations of motion the Schr\"odinger equation for this, or any other, system is strictly linear. We find that solutions to the Schr\"odinger equation for this system, even for moderate values of $\beta$, delocalise so rapidly that obtaining accurate solutions is not possible for us. This does, however, reinforce the lack of correspondence between classical dynamics and Schr\"odinger evolution for this system.

Following past work~\cite{Car93,Gis93,Gis93b,Heg93,Wis96,Ple98,spi1,bru1,bru2,zur1,sch1}  on recovering classically chaotic like orbits from a system's quantum counterpart we solve the unravelling of the master equation~(\ref{eq:qsd}) with Hamiltonian~(\ref{eq:Ham}). For this example there are three points of note with regard to possible choices of the environmental degrees of freedom. Firstly, coupling to an environment helps localise the system's state vector and hence produce a well defined, classical-like, trajectory. Secondly, as the classical system is Hamiltonian and therefore conservative, we must chose the environment of each oscillator so that energy exchange is minimised  between any part of the system and the environmental degrees of freedom. Thirdly, we should specify a physically reasonable environment.

\begin{figure}[!t]
\begin{center}
\resizebox*{0.48\textwidth}{!}{\includegraphics{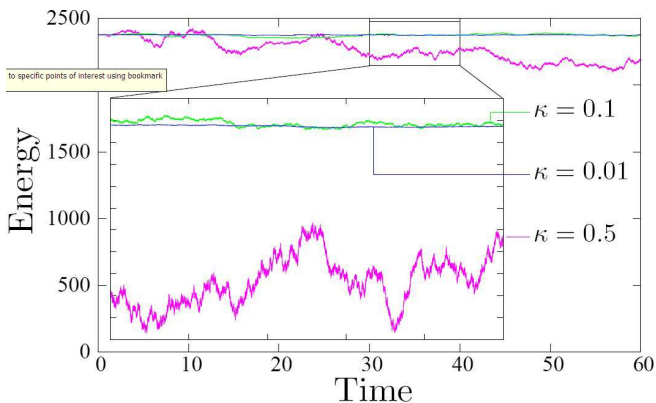}}
\caption{(colour on-line) Total system energy, with magnified section inset, computed by substituting $\EX{q_i}$ and $\EX{p_i}$ into the Hamiltonian~(\ref{eq:Ham}) for three different couplings ($\kappa$) to the environment. Note, all quantities are dimensionless.
\label{fig:2}}
\end{center}
\end{figure}

\begin{figure}[!t]
\begin{center}
\resizebox*{0.48\textwidth}{!}{\includegraphics{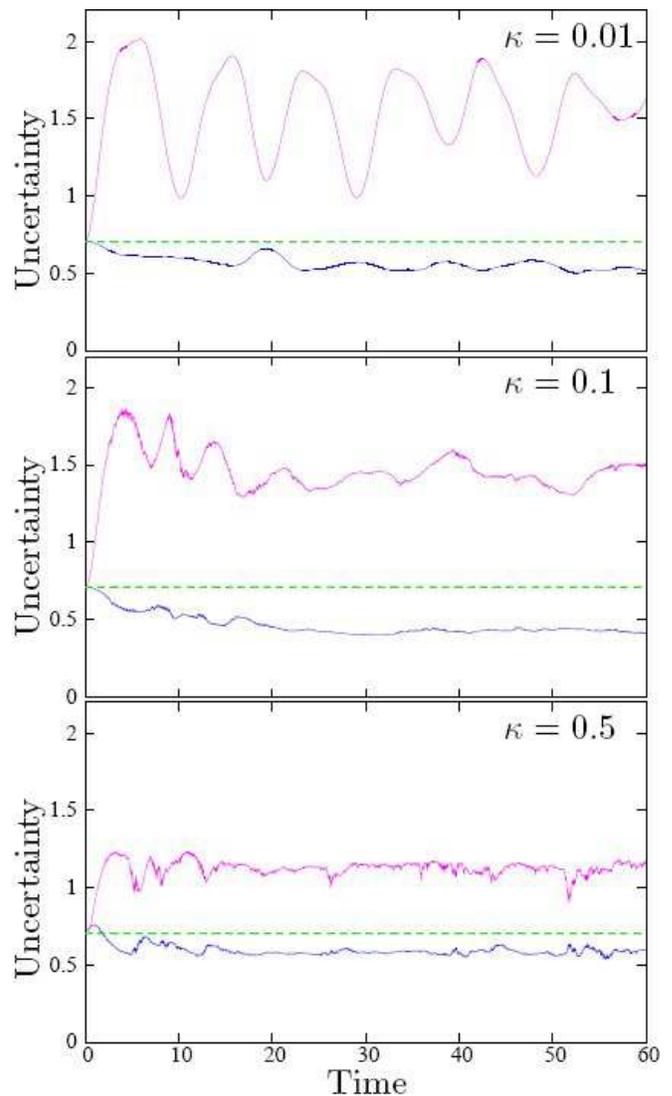}}
\caption{(colour on-line) Uncertainty in position (light grey/magenta) and momentum (dark grey/blue) as a function of time for the first component for three different couplings ($\kappa$) to the environment. Values beneath the dashed (green) line indicate squeezing. Note, all quantities are dimensionless.
\label{fig:un}}
\end{center}
\end{figure}

\begin{figure}[!t]
\begin{center}
\resizebox*{0.48\textwidth}{!}{\includegraphics{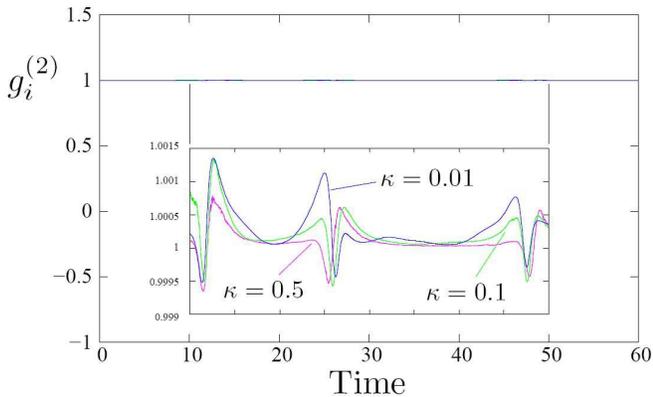}}
\caption{$g_i^{(2)}$ $(i=1)$ coefficient, with magnified section inset, as a function of time for three different couplings ($\kappa$) to the environment. Classical-like motion yields $g_i^{(2)} = 1$ as this implies a Poissonian statistics for the state of the system. Note, all quantities are dimensionless.
\label{fig:g2}}
\end{center}
\end{figure}

In this work we have chosen initially one of the most obvious candidates for the environment which satisfies all these conditions. Explicitly we have set each Lindblad $L_i=\kappa q_i,\ i=1,2,3$ corresponding to the continuous measurement of position. This unravelling also corresponds to that of the  Master equation for a weakly coupled, high temperature, thermal environment~\cite{bha1}. Here $\kappa$ represents the magnitude of the coupling between each component of the system and its respective environment. In this work we use several values of $\kappa$. In figure~\ref{fig:1}(b) we use an intermediate coupling ($\kappa=0.1$) whilst in figures~\ref{fig:new2a},~\ref{fig:2},~\ref{fig:un} and~\ref{fig:g2} we also present results for weak ($\kappa=0.01$) and strong ($\kappa=0.5$) couplings.

As our initial  state, and for the best possible match with the classical initial conditions, we chose a tensor product of coherent states for which the quantum expectation values in position and momentum  are centred in $q-p$ phase  plane at $q_1=-0.2/\beta, q_2=0.05/\beta, q_3=0.15/\beta$ and $p_i=0$ where $i=1,2,3$. Alternatively, we can express this initial condition explicitly as translated vacuum states by $D_1(-0.2/\beta)D_2(0.05/\beta)D_3(-0.15/\beta)\ket{000}$ where $D_i(.)$ is the displacement operator in position for each component of the system. Here we have chosen $\beta=1/2000$ as this is the smallest value for which we can solve~(\ref{eq:qsd}) both accurately and within a reasonable time frame. In order to help the reader quantify the time scale over which our results are presented we note that the log time associated with our chosen value of beta is $\log (1/\beta)\approx 3.3$~\cite{Steck02,Zurek94}. This is much shorter than the period over which we present the evolution of the system's trajectories.

In figure~\ref{fig:1}(b) we show the dynamics of the quantum expectation values of the position and momentum operators for each oscillator. These have been scaled by a factor of $\beta$ so that they may be compared with figure~\ref{fig:1}(a). Here we see very good agreement initially, and similar characteristics throughout, the displayed dynamics. Indeed the trajectories are similar enough that it is impossible to determine from the graph alone which plot shows the classical and which the quantum evolution. We note that these curves begin to differ  after a short period of time. However, this is not unexpected as the system we are analysing is chaotic.
In order to make the reasonable comparison of these results readily available we also include a graph of the evolution of $q_1$ and $\EX{q_1}$ as a function of time in Fig.~\ref{fig:new2a} for three different couplings to the environment. It is apparent that there is a very good match between the quantum expectation values and the classical trajectory for $\kappa=0.01$.

We note that simply by including an environment our system no longer undergoes Hamiltonian evolution. In other words, in order to be able to recover classical like trajectories of quantum systems whose classical counterparts exhibit Hamiltonian chaos we include environmental degrees of freedom that imply non-Hamiltonian evolution of the quantum system. However, using a sufficiently low coupling strength to the environment results in a concomitant reduction both in energy exchange between the system and it's environment and the localisation of the state vector.
We now verify that the solutions to equation~(\ref{eq:qsd}) are to, good approximation, conservative.  This does indeed appear to be the case for both intermediate and weak couplings ($\kappa = 0.1$ and $0.01$) but not for the stronger coupling ($\kappa = 0.5$). This can be seen in figure~\ref{fig:2} where we show the total energy found by substituting $\EX{q_i}$ and $\EX{p_i}$ for ${q_i}$ and ${p_i}$ into the Hamiltonian~(\ref{eq:Ham}), i.e.
\begin{equation}
\mathrm{Energy} = \sum_i \frac{1}{2}\EX{p_i}^2+\frac{\beta^2}{32}\EX{q_i}^4+
\frac{\beta^2}{2}\sum_{i \neq j} \EX{q_i}^2\EX{q_j}^2
\end{equation}
where $i,j=1,2,3$. We note that we do not compute $\EX{H}$ as we wish to compare directly with the equivalent classical calculation.

Next we verify localisation of the state vector by computing the uncertainty in position and momentum for the first oscillator for three different couplings to the environment. Because both $\Delta q_i=\sqrt{\EX{q_i^2}-\EX{q_i}^2}$ and $\Delta p_i=\sqrt{\EX{p_i^2}-\EX{p_i}^2}$ between components behave in a similar fashion, we do not show results for the other two oscillators here. As is evident from figure~\ref{fig:un}  the  interaction with the environment causes system's state vector to localise within each of the component spaces. It is also apparent from this figure that the level of localisation is dependent of the coupling between each of the system's components and their respective environments. We also note that as the system evolves it's states become squeezed in each of the momentum variables. Unlike the results presented in figure~\ref{fig:1}(b) we have not scaled these uncertainty values by $\beta=1/2000$. Consequently, for direct comparison with the first two figures the results presented in figure~\ref{fig:un} should be divided by $2000$. Hence, the uncertainty in either position or momentum can be seen to be quite negligible when compared with the trajectory of quantum expectation values plotted in figure~\ref{fig:1}(b).

We can extract further information on the dynamics of this system simply by borrowing a technique from quantum optics. Namely through analysing the photon statistics (bunching of photons) described by the second order
correlations~\cite{ScullyZ97,plk}
\begin{equation}
g_{i}^{(2)}=\frac{\left\langle n_{i}^{2}\right\rangle -\left\langle
n_{i}\right\rangle }{\left\langle n_{i}\right\rangle ^{2}},\ i=1,2,3\label{eq:g2}%
\end{equation}
Where $n$ is the number operator. 
Values of $g^{\left(  2\right)  }$ greater than 1 indicate photon bunching
 where  photons arrive in groups while values of $g^{\left(
2\right)  }$ smaller than one indicate antibunching, a purely quantum mechanical phenomena representing precisely regular arrival
of photons. However $g^{\left(  2\right)  }=1$ corresponds to Poissonian statistics and which is what we would expect from the state of our system should it be undergoing a classical like evolution. As we can see from figure~\ref{fig:g2} this is indeed the case (where, again, we have only shown data for the first component).

\begin{figure}[!t]
\begin{center}
\resizebox*{0.48\textwidth}{!}{\includegraphics{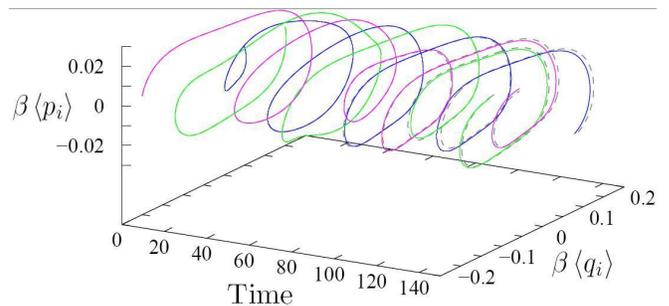}}
\caption{(colour on-line) An example chaotic-like trajectory for the scaled quantum expectation values $\beta\EX{q_i}$ and $\beta\EX{p_i}$, for weak measurement of the first oscillator only. Here $L_1=\kappa q_1, L_2=L_3=0$ for $\kappa=0.01$ and $\beta=1/2000$.  Quantum trajectories for oscillator one are shown in magenta (medium grey), for two in blue (dark grey) and three in green (light grey). The corresponding classical dynamics are shown with a dashed grey line. Note, all quantities are dimensionless.
\label{fig:5}}
\end{center}
\end{figure}

In the  discussion above we have considered what happens when  we perform  simultaneous, continuous, measurements of the position of each of the components of the system. As the weak measurement limit is approached we find good agreement between classical and quantum dynamics and the system becomes, to good approximation, conservative.
We  now demonstrate that there exists a weaker condition under which we can produce the same outcome. That is, we show that weak measurement of only one of the position variables is sufficient produce, to very good approximation, classical-like
trajectories. As no one oscillator has a privileged status over the others we set, without loss of generality, $L_1=\kappa q_1, L_2=L_3=0$ with $\kappa=0.01$ and $\beta=1/2000$. Again we solve the unravelling of the master equation~(\ref{eq:qsd}) with Hamiltonian~(\ref{eq:Ham}) and the initial state $D_1(-0.2/\beta)D_2(0.05/\beta)D_3(-0.15/\beta)\ket{000}$. As we can see from figure~\ref{fig:5} even under these conditions we recover quantum trajectories whose expectation values match very well indeed with those of the equivalent classical system. 

\section{Conclusion}
For any given $\beta$ and an initially localised state there will be some agreement between the dynamics of the classical system and the evolution of the quantum expectation values of the corresponding quantum operators. However, after a short period of time the quantum state vector will begin to delocalise and differences between the predictions of each theory become apparent. We have demonstrated, by localising the state vector through  measurement of the position of one or all components of the system, that near classical like dynamics can be recovered through an unravelling the master equation. For this work we have chosen quantum state diffusion. However, it is likely that any other unravelling will produce similar results. Such detailed analysis is beyond the scope of this work and would belong in a more in depth study. We also note, as a subject for future study, that it would be interesting to determine the conditions under which measurement of a subset of the degrees of freedom of an $N$-body system results in the localisation of the state vector.

Finally we would like to observe that following~\cite{everitt05} it would be interesting to characterise the entanglement between the components of this system. As it may well be the case that for this example, as well as the one studied in~\cite{everitt05}, that the entanglement does not necessarily vanish in the classical limit.  Unfortunately current restrictions on computational power prevent us from conducting this study at the present time.
However, from the last result presented here we intuitively feel that there must persistently exist at least a small degree of entanglement between the first component and each of the other two. In order to justify this statement we propose the following argument. First, consider the  extension to the tensor product space of the Lindblad operator, explicitly this is $\kappa q_1 \otimes 1_2 \otimes 1_3$. Now, by examining equation~(\ref{eq:qsd}) we see that if the state of the system was separable, the last two terms (those responsible for the localisation of $\ket{\psi}$) would not affect the components of the state vector for the second and third degrees of freedom. As introduction of this environmental degree of freedom localises the state vector and result in the recovery of a classical like trajectory, we therefore propose that the state vector must posses some non-zero entanglement.

\begin{acknowledgments}
The author would like to thank Professor~F.V.~Kusmartsev, Professor~S.A.~Alexandrov, and Dr~J.H.~Samson whose stimulating discussion with the author regarding~\cite{everitt05,everitt05b} and systems that exhibit Hamiltonian chaos whilst visiting Loughborough University originated this work. The author would also like to thank Dr~J.F.~Ralph, Prof~T.D.~Clark and Dr~T.P.~Spiller for interesting and informative discussions.
\end{acknowledgments}

\bibliography{references}

\end{document}